\documentstyle[aps,prb]{revtex}
\include{psfig}
\begin{document}

\title{Resonant multiple Andreev reflections in mesoscopic superconducting
junctions} \author{G. Johansson$^{1}$, E. N. Bratus$^{1,2}$, V.S.
Shumeiko$^{1,2}$, and G. Wendin $^{1}$}
\address{$^{1}$ Department of
Microelectronics and Nanoscience, Chalmers University of Technology and
G\"oteborg University, S-412
96 G\"oteborg, Sweden \\ $^{2}$ B. Verkin Institute of Low Temperature
Physics and Engineering, 310164 Kharkov, Ukraine} 

\date{\today}
\maketitle
\draft

\begin{abstract}
We investigate the properties of subharmonic gap structure (SGS) in
superconducting quantum contacts with normal-electron resonances. We find
two distinct new features of the SGS in resonant junctions which
distinguish them from non-resonant point contacts:
(i) The odd-order structures on the current-voltage characteristics of
resonant junctions are strongly enhanced and have pronounced peaks, while
the even-order structures are suppressed, in the case of a normal electron
resonance being close to the Fermi level.
(ii) Tremendous current peaks develop at $eV=\pm 2E_0$ where $E_0$ indicates
a distance of the resonance to the Fermi level. These properties are
determined by the effect of narrowing of the resonance
during multiple Andreev reflections and by overlap of electron and hole
resonances. 
\end{abstract}

\twocolumn

\section{Introduction}
Resonant tunneling often plays an important role in current transport in
transmissive superconducting junctions. The presence of impurities in the
tunnel barriers may strongly affect Josephson tunneling 
\cite{Asl2,Tart,Gla,Dev} as well as Andreev transport under applied
voltage \cite{Gla2,Golub} due to the resonant tunneling through localized 
impurity levels. A mechanism of resonant tunneling was assumed to be
responsible for the unusual properties of junctions with disordered
semiconducting barriers \cite{Ovi} and grain boundary Josephson junctions
of high Tc materials \cite{Gross}. Furthermore, in clean superconductor
-semiconductor
junctions, mobile electrons confined between the Schottky barriers form
resonant states which determine specific properties of such junctions 
\cite{Asl,She}. Similar resonant states are also 
important in recently fabricated ballistic superconductor-2D electron gas
(2DEG) structures \cite{Taka} where they are formed by the electron
reflections by the gate potential. Moreover, the possibility to create 
quantum point contacts and quantum dots in such structures allows one to 
investigate quantum resonant transport through well resolved resonant levels
\cite{Yacobi}. Quantum
resonant transport was also observed in metallic dots \cite{Ralph}, and in
contacts containing single wall nanotubes \cite{Delft1,Rice}.

The properties of dc Josephson current in quantum resonant junctions as 
well as Andreev quantization have been discussed in various publications
\cite{Volk,Bee,Cre,WSh}. The subharmonic gap structure (SGS) in resonant
junctions and the effect of the resonance on multiple
Andreev reflections (MAR) is less investigated.
Meanwhile, detailed theory of SGS in quantum point contacts 
\cite{Sh1,Sh2,Ave1,Ye1,BB} in combination with precise experiments on
controllable break junctions \cite{Jan,Urb} was found to be a powerful tool
for investigation of intrinsic properties of the atomic-size contacts. 
Our preliminary results \cite{Joh} have shown that SGS in resonant junction 
drastically differs from the SGS in non-resonant junction; similar results 
were obtained by different methods in Ref. \cite{Ye2}.  In
this paper, we present a detailed study of interplay of MAR with
Breit-Wigner resonances in quantum junctions.

The structure of the paper is the following. In Section II we derive
equations for the inelastic scattering amplitudes in resonant junctions. 
In Section
III we discuss properties of the normal electron resonance in the proximity
region between the superconducting electrodes. Results of numerical
calculations of the current-voltage characteristic for different
positions and widths of the resonance are presented in Section IV. Section
V is devoted to perturbative analysis of resonant SGS.

\section{Scattering amplitudes.}

We will consider a junction consisting of a ballistic normally conducting
region separated from the superconducting electrodes by tunnel barriers,
Fig. 1. The length of the junction $L$ is assumed to be smaller than the
coherence length, and therefore the distance between normal resonances will
exceed the superconducting energy gap,
$v_{F}/L>\Delta$ ($v_{F}$ is the normal electron Fermi velocity,
$\hbar=1$). We will also assume that the resonances are well separated,
$\Gamma\ll v_{F}/L$, where $\Gamma$ is the resonance half-width, and that 
Coulomb charging effects do not dominate in the subgap voltage region,
$E_C<2\Delta$, where $E_C$ is a Coulomb gap.

We will apply Landauer-B\"uttiker scattering theory \cite{Lan,But,Imry}
extended to superconducting junctions \cite{Sh2} for calculating the
current. In voltage biased superconductive junctions, the quasiparticle
scattering is inelastic due to time dependence of the superconducting phase
difference across the junction, $\phi(t)=2eVt$, and the scattering state
wave functions consist of linear combinations of harmonics (sidebands) with
energies $E_n=E+neV$ shifted by integer number of quanta $eV$ with respect
to the energy $E$ of the incoming wave. Below we will consider one
transport mode in the junction and choose the scattering state wave
functions in the left ($L$) and right ($R$) electrodes having the form
\begin{mathletters}
\label{psi}
\begin{eqnarray}
\label{psiL}
\psi_L &=&
e^{-iEt} \left[
\delta_{j1}u_0^+e^{ik_0^+x}+
\delta_{j2}u_0^-e^{-ik_0^-x}\right]+ \nonumber \\
&&\sum_{n=-\infty}^{\infty} e^{-iE_nt} \left[a_nu_n^-e^{ik_n^-x}+
c_nu_n^+e^{-ik_n^+x}\right] \\
\label{psiR}
\psi_R &=&
e^{i\sigma_z \phi(t)}
\left\{
e^{-iEt} \left[
\delta_{j3}u_0^+e^{-ik_0^+x}+
\delta_{j4}u_0^-e^{ik_0^-x}\right] \right. + \nonumber \\
&& \left. \sum_{n=-\infty}^{\infty} e^{-iE_nt}\left[b_nu_n^-e^{-ik_n^-x}+
f_nu_n^+e^{ik_n^+x} \right]
\right\},
\end{eqnarray}
\end{mathletters}
where $u_n^\pm$ are (non-normalized) two-component elementary solutions of
the Bogoliubov-de Gennes equations,
\begin{equation}
u_n^\pm={1\over \sqrt 2}
\left(
\begin{array}{c}
e^{\pm\gamma_n/2}\\
\sigma_n e^{\mp\gamma_n/2}
\end{array}
\right).
\label{u}
\end{equation}
In this equation
$$
e^{\gamma_n}={|E_n|+\xi_n\over\Delta},\;\; \xi_n=
\left\{
\begin{array}{lr}
\sqrt{E^2_n-\Delta^2}, & |E_n|>\Delta\cr i\sigma_n\sqrt{\Delta^2-E^2_n},
&
|E_n|<\Delta \end{array}\right. ,
$$
$$
\sigma_n=\mbox{sgn} (E_n),\;\;\;
k_n^\pm=\sqrt{2m(E_F\pm \sigma_n\xi_n)}.
$$
Index $j=1-4$ in Eqs. (\ref{psi}) labels scattering states of the electron-
and hole-like quasiparticles incoming from the left ($j=1,2$) and right
($j=3,4$).
The form of wave functions in Eqs. (\ref{psi}), (\ref{u}) assumes that
superconducting electrodes serve as equilibrium quasiparticle reservoirs,
and that the potential difference between the reservoirs is absorbed into
the time-dependent factor $e^{i\phi(t)}$ in Eq. (\ref{psiR}) due to
appropriate choice of the gauge.

To match the wave functions in Eqs. (\ref{psi}) we will apply a transfer
matrix technique. In the present case of an inelastic scattering problem,
the connection between $\psi_L$ and $\psi_R$ is non-local in time, and the
corresponding transfer matrix ${\bf T}^S_{nm}$ mixes the sidebands,
\begin{equation}
{A\choose B}_{Ln}= \sum_m {\bf T}^S_{nm}{A\choose B}_{Rm} . 
\label{matchS}
\end{equation}
The matrix ${\bf T}^S_{nm}$ is a $4\times 4$ matrix defined on a space of
wave function coefficients, $A=(A^+,A^-)$, $B=(B^+,B^-)$,
\begin{eqnarray}
\psi_{n}&=&e^{-iE_nt}\left[A^+_nu_n^+e^{ik_n^+x}+A^-_nu_n^+e^{-ik_n^+x}
\right. \nonumber \\
&+& \left. B^+_n u_n^-e^{ik_n^-x}+B^-_nu_n^-e^{-ik_n^-x}\right] .
\label{psiLn}
\end{eqnarray}
The transfer matrix in Eq. (\ref{matchS}) can be expressed, similarly to the
case of unbiased junctions \cite{WSh}, through a transfer matrix $T(E)$
associated with elastic electron scattering by the normal junction.
Let us introduce auxiliary normal regions between the superconductors and
the junction, the length of which normal regions will be put equal to zero
at the end of the calculations. The wave function in the normal region
has the form
\begin{equation}
\psi_{n}^N=
{A_n^{N+}e^{ik_n^{N+}x}+A_n^{N-}e^{-ik_n^{N+}x}\choose B_n^{N+}
e^{ik_n^{N-}x} + B_n^{N-}e^{-ik_n^{N-}x}}e^{-iE_nt},
\label{psiLnN}
\end{equation}
where $k_n^{N\pm}=\sqrt{2m(E_F\pm E_n)}$ is the normal electron wave
vector. The wave functions Eq. (\ref{psiLnN}) in the left and right normal
regions are related as
\begin{equation}
{{ A}^N\choose { B}^N}_{Ln} ={\bf T}^{N}_n{{ A}^N\choose{ B}^N}_{Rn},
\;\;\;{\bf T}^{N}_n=\left( \begin{array}{cc} T(E_n)&0\\ 0&T(-E_n)\\
\end{array} \right).
\label{matchT}
\end{equation}
We note that the transfer matrix ${ T}(E)$ describes scattering of the
normal electrons by the actual potential of the junctions at a given
voltage, i.e. it includes effects of potential deformation due to applied
voltage, $T(E;V)$.

Continuous matching at the left SN interface yields in quasiclassical
approximation $k_n\approx k_n^N\approx k_F$, %
\begin{equation}
{A^N\choose B^N}_{Ln} ={\bf T}^{NS}_n{ A\choose B}_{Ln},\;\; {\bf
T}^{NS}_n=\left( \begin{array}{cc}
e^{\gamma_n/2}&e^{-\gamma_n/2}\\
\sigma_n e^{-\gamma_n/2} & \sigma_n e^{\gamma_n/2}\\ \end{array} \right).
\label{matchL}
\end{equation}
A matching condition at the right NS interface is derived in a similar way
but an additional time-dependent factor $e^{i\sigma_z eVt}$ in Eq.
(\ref{psiR}) must be taken into account. The latter gives different
equations for upper (electron) and lower (hole) components of the
coefficient vectors:
\begin{equation}
{ A}_{Rn} ^N={\bf P}^+{\bf T}^{NS}_{n+1}{ A\choose B}_{R(n+1)},\;\;
{ B}_{Rn}^N={\bf P}^-{\bf T}^{NS}_{n-1}{ A\choose B}_{R(n-1)}.
\label{matchR}
\end{equation}
In this equation, ${\bf P}^\pm$ are projectors on upper/lower vector
components.

Combination of Eqs. (\ref{matchT})-(\ref{matchR}) gives the following
equation for the transfer matrix in Eq. (\ref{matchS}) 
\begin{equation}
{\bf T}^S_{nm}=\sum_\pm
({\bf T}^{NS}_{n})^{-1}{\bf T}^{N}_{n}
{\bf P}^\pm{\bf T}^{NS}_{m}\delta_{m(n\pm 1)}.
\label{match}
\end{equation}
Normal electron transfer matrix enters this equation with different
arguments $\pm E_n$. This energy difference introduces effects of
electron-hole dephasing during quasiparticle propagation through the
junction. In non-resonant short constrictions, the energy dispersion of the
transfer matrix is negligible, and equation (\ref{match}) is equivalent to
the matching equation derived in Ref.\cite{Sh2}. In resonant junctions (and
also in long SNS and SIS junctions \cite{WSh2}) dephasing effects are
important.

The matching equations (\ref{matchS}),and (\ref{match}) can be written
in an equivalent form,
\begin{equation}
{\bf P}^\pm{\bf T}^{NS}_{n}{ A\choose B}_{Ln}= T(\pm E_n){\bf P}^\pm{\bf
T}^{NS}_{Rn\pm 1}{ A\choose B}_{R(n\pm 1)}.
\label{match1}
\end{equation}
Applied to the scattering state wave functions in Eqs. (\ref{psi}), it
yields the following recurrences for the scattering amplitudes:
\begin{mathletters}
\label{mar}
\begin{eqnarray}
&e^{\sigma_z\gamma/2}
\delta_{n0} {\delta_{j1}\choose \delta_{j2}}+
e^{-\sigma_z\gamma_n/2}{a\choose c}_n= \nonumber \\
&{T}(E_n)
\left[ e^{\sigma_z\gamma/2}
\delta_{(n+1)0} {\delta_{j3}\choose \delta_{j4}}+
e^{-\sigma_z\gamma_{n+1}/2}{f\choose b}_{n+1} \right] \\
\label{up}
&e^{-\sigma_z\gamma/2}
\delta_{n0} {\delta_{j1}\choose \delta_{j2}}+
e^{\sigma_z\gamma_n/2}{a\choose c}_n= \nonumber \\
&{ T}(-E_n)\sigma_n\sigma_{n-1}
\left[ e^{-\sigma_z\gamma/2}
\delta_{(n-1)0} {\delta_{j3}\choose \delta_{j4}}+
e^{\sigma_z\gamma_{n-1}/2}{f\choose b}_{n-1} \right].
\label{down}
\end{eqnarray}
\end{mathletters}
Analytical
solutions of the recurrences in Eqs. (\ref{mar}) can be presented in
chain-fraction form (see Appendix A) similar to the case of non-resonant
junctions \cite{Sh1,Sh2}.

\section{Model for resonance.}

Now we will specify the transfer matrix for the resonant junction. We will
restrict ourselves to symmetric junctions,
$T_{11}=T_{22}^\ast=1/d$ and $T_{21}=T_{12}^\ast=r/d$ and assume the
Breit-Wigner resonance form for transmission and reflection amplitudes $d$
and $r$ respectively,
\begin{equation}
d(E)={i\Gamma\over E-E_r+i\Gamma},\;\;r(E)=-{E-E_r\over E-E_r+i\Gamma}.
\label{bw}
\end{equation}
The position of the resonance level $E_r$ as well as the resonance
half-width $\Gamma$ are generally dependent on the applied voltage.
However, while the subharmonic gap structure is affected in an essential
way by the position of the resonance, the dependence on the resonance width
is less important. Thus we will assume $\Gamma=const$. We will not specify
the voltage dependence of resonance level position, but rather present the
current as a function of two variables: driving voltage and resonance
position, $I(V,E_r)$. The current voltage characteristics can then be
reconstructed from such a dependence by specifying the $E_r(V)$ dependence
determined by self-consistent distribution of the electric potential across
the junction.

The normal electron resonance, being confined between superconducting
electrodes, possesses specific properties which will be important for
further analysis of the resonant MAR. Since the transfer matrix $T(E)$
enters the recurrences for the scattering amplitudes in Eq. (\ref{mar}) at
two different energies $\pm E$, the proximity resonance consists of two,
electron and hole, resonances situated symmetrically with respect to the
Fermi level, $E=\pm E_r$.
Within the adopted approach, the current is calculated by using the
scattering amplitudes defined in the {\em superconducting electrodes} [see
further Eq. (\ref{I})] and the recurrences in Eq. (\ref{mar}) are
formulated for these amplitudes. Although equivalent, such an approach is
different from the discussion of MAR amplitudes in the normal region of the
junction (see, e.g. \cite{KBT}). Within our approach, the
non-superconducting region of the junction is considered as a black box and
is represented by the transfer matrix $T(E)$. Due to the different choices
of gauge in the left and right electrodes, the resonance is seen from the
left and right electrodes at different energies [cf. Eqs. (\ref{matchL})
and (\ref{matchR})]. Indeed, the resonances are seen from the left
electrode at $E=\pm E_r$, i.e. quasiparticles
incoming from the left undergo resonant transition if $E=\pm E_r$, while the
resonances are seen from the right electrode at $E=\pm (E_r+eV)$, as shown
in Fig. 2. In the scattering diagram in Fig. 2c, the resonance
therefore is presented with two segments:  $E_n=E_r \leftrightarrow
E_{n+1}=E_r+eV$ for the electron resonance, and $E_n=-E_r \leftrightarrow
E_{n-1}=-E_r-eV$ for the hole resonance.

There is a symmetry between the scattering states originating from the left
and right electrodes:
\begin{equation}
{a\choose c}_{n,3}(\gamma,\Gamma, E_0)=\sigma_0\sigma_n {f\choose
b}_{n,1}(-\gamma, -\Gamma, -E_0), \label{sym13}
\end{equation}
with an analogous relation for the second pair of scattering amplitudes. In
Eq. (\ref{sym13}), $E_0=E_r+eV/2$
is the distance of the normal resonance level with respect to the midpoint
between the chemical potentials in the left and right electrodes. Equation
(\ref{sym13}) leads to a symmetry property of the current which is an even
function of the resonance position $E_0$: $I(V,E_0)=I(V,-E_0)$.
Below we will indicate the resonance position by means of the energy $E_0$
and abbreviate the Breit-Wigner amplitudes (\ref{bw}),

\begin{equation}
d_n^{\pm}={i\Gamma\over E_n^{\pm}+i\Gamma},\;\; r_n^{\pm}={E_n^{\pm}\over
E_n^{\pm}+i\Gamma}, \;\;\; E_n^{\pm}=E_n\mp (E_0-eV/2).
\end{equation}
%

\section{dc Current.}

In the quasiclassical approximation, the equation for the current reads
\cite{Sh1,Sh2}
\begin{eqnarray}
&I= {e\Delta\over2\pi}\int^{\infty}_{\Delta} {dE\over \xi}
\sum_{n=odd} \cosh(Re \gamma_n)\tanh{E\over 2T} \nonumber \\ 
&\left[
\sum_{j=1,2}\left(|b_{n,j}|^2-|f_{n,j}|^2\right)-
\sum_{j=3,4}\left(|c_{n,j}|^2-|a_{n,j}|^2\right)\right]
\label{I}
\end{eqnarray}
The current in Eq. (\ref{I}) is calculated using transmitted states (in the
right and left electrodes for scattering states $j=1,2$ and $j=3,4$
respectively), and it consists of contributions from all odd sidebands. By
virtue of the symmetry equations 
\begin{eqnarray}
{f\choose b}_{n,2}(\gamma, { T}) & = &
{b\choose f}_{n,1}(-\gamma, { T^\ast}) , \nonumber \\
{a\choose c}_{n,2}(\gamma, {T})
& = & {c\choose a}_{n,1}(-\gamma, { T^\ast}) , 
\label{sym12}
\end{eqnarray}
directly following from Eqs. (\ref{mar}) (analogous relations hold for the 
scattering states $j=3,4$) and symmetry equations (\ref{sym13}), the
current in Eq. (\ref{I}) can be expressed through the sideband contributions
\begin{equation}
K_{n}=\left[|b_{n}|^2-|f_{n}|^2\right]\cosh(Re \gamma_n)
\label{K}
\end{equation}
of one single scattering state ($j=1$, index $j$ is omitted), 
\begin{equation}
I= {e\Delta\over2\pi}\int^{\infty}_{\Delta} {dE\over \xi}
\sum_{n=odd}[K_{n}-\bar K_{n} + (E_0\rightarrow -E_0)]\tanh{E\over 2T}, 
\label{I'}
\end{equation}
where $\bar K_{n}=K_{n}(-\xi_n, -\Gamma)$. 

Equation Eq. (\ref{I'}) together
with the recurrences in Eqs. (\ref{mar}) provide a basis for
numerical calculation of the current. The calculation of scattering 
amplitudes
should obey the boundary condition at $\pm$ infinity where the amplitudes
approach zero. The simplest way to obtain such solutions is to iterate the
recurrences from large $|E_{n}|$ towards $E$. The correct
solution will then grow exponentially and numerically "kill" the solution 
growing at infinity. By this procedure one gets the correct scattering 
states for each incoming quasiparticle at every energy.

The results of numerical calculation of current-voltage characteristics (IVC) 
are presented  in Fig. 3 for
different values of resonance level position $E_0=const$. This particular
case corresponds to a perfectly symmetric distribution of the electric
potential across the junction with $E_0$ indicating the departure of the
resonance level from the Fermi level in equilibrium $(V=0)$. The IVC with
the resonance level situated at the Fermi level, $E_0=0$, shows an onset of
the single-particle current at $eV=2\Delta$ accompanied by a current peak
caused by large density of states near the superconducting gap [see below
Eqs. (\ref{I1res}-\ref{I1asym})].
Such behaviour of the single-particle current has been observed in the
experiments on metallic dots \cite{Ralph} and carbon-nanotube junctions
\cite{Dekker}. A striking feature of this IVC is the absence of current
structure at $eV=\Delta$, while the structure at $eV=2\Delta/3$ is
pronounced, consisting of a peak similar to the structure of the
single-particle current. Calculation of the IVC at lower voltage,
presented in Fig. 4, shows the same feature - only odd subharmonic gap
structures are present.

If the resonant level departs from the Fermi level and $E_0\neq 0$, the
single-particle current onset shifts towards larger voltage, $eV>2\Delta$,
and the current peak broadens. A striking feature in this case is the
development of a huge current peak at voltages lower than the position of
the structure of single-particle current. This peak, associated with
resonant pair current (see below Sec. V), appears as soon as
$E_0>\Delta/2$ and is situated at voltage $eV=2E_0$ which coincides with
position of the resonant current onset in normal junctions. 
If the resonance level departs far from the Fermi level,
$E_0\gg\Delta$, the IVCs in the subgap voltage region $eV<2\Delta$ 
approach the form typical for non-resonant point
contacts, as could be expected, while a broadening of the resonance,
$\Gamma\gg \Delta$, gives rise to SNS-type IVCs, as shown in Fig. 5.

A complete description of the current in resonant junctions is given by the
function $I(V,E_0)$, as already mentioned in Section III. A plot of this
function is presented in Fig. 6. The IVCs plotted in Figs. 3-5 correspond
to horizontal cuts ($E_0=const$) of the plot in Fig. 6. In Fig. 6a, the
light wedge-like region at $eV>2\Delta$ corresponds to the resonance
single-particle current. The resonant peak of the pair current is seen as the
light streaks directed along the lines $E_0=\pm eV/2$, the structure
starts at $eV=\Delta$. 
Fig. 6b presents a similar plot for the region of small voltage,
$eV<\Delta$. The picture shows quite complex structure of the current
consisting of wedge-like plateaux of the resonant current as well as of
light streaks corresponding to current peaks.

In order to interpret the features of the IVCs one needs to analyze the
properties of the side-band currents $K_n$ presented in Eqs. (\ref{K}) and
(\ref{Kexp}).

\section{Discussion}

A convenient expression for analysis of the subharmonic gap structure is
derived in Appendix B:
$$I_{SGS}(V, E_0)=\sum_{n=1}^\infty I_n(V, E_0),\\ $$
\begin{eqnarray}
I_n(V, E_0)={e\Delta\over2\pi}\int^{neV-\Delta}_{\Delta}
\tanh{E\over 2T}{dE\over\xi} \nonumber \\
\left[ \tilde K_{-n} - \tilde{\bar K}_{-n} + (E_0\rightarrow -E_0)\right]
\label{In}
\end{eqnarray}
In Eq. (\ref{In}) only contributions of processes of creation
of real excitations (transitions across the gap $E>\Delta\rightarrow
E_n<-\Delta$) are retained which are responsible for the subharmonic gap 
structure \cite{Sh2} while a contribution of thermal excitations is omitted. 
Furthermore, the sum
over the sideband currents in Eq. (\ref{I'}) is now rearranged in order
to explicitly separate the contributions of all inelastic channels
[contributions of {\em even} inelastic channels are hidden in Eq.
(\ref{I'})]. This is done by proper renormalization of the sideband
currents $K_n\rightarrow \tilde K_{n}$ presented in Appendix B, the 
equation for $\tilde K_{n}$
being given in Eq. (\ref{Ktilde}). We will now develop a perturbative
analysis of the current in the limit of small width of the resonance,
$\Gamma\ll\Delta$, and zero temperature.

\subsection{Single-particle current}

The single-particle current is given by the first term in Eq. (\ref{In}).
In accordance  with Eq. (\ref{Ktilde}), it has explicit form
\begin{eqnarray}
I_1={4e \over \pi}
\int_{\Delta}^{eV-\Delta} dE\,
{|E_{-1}|\xi\xi_{-1}\over \Delta^3} \nonumber \\
\left\{
D_0^-
\left(
{e^{-\gamma}\over P_1}+{e^{\gamma}\over \bar P_1} \right)
+
\left(
E_{0}\rightarrow -E_{0}
\right)
\right\},
\label{I1}
\end{eqnarray}
$P_1$ is defined in Eq. (\ref{P'}).
This current has no contribution from Andreev reflections and it has only
one resonance. It is sufficient to consider only scattering states incoming
from the left [the first term in curly brackets in Eq. (\ref{I1})], 
the resonance equation in this case being  $E_0^-=0$. 
The resonance is only involved if it belongs to the integration interval.
This determines the resonance region $eV/2>\Delta+|E_{0}|$ in the plane
$(V,E_0)$ (region $I$ in Fig. 7). The resonant scattering diagram is shown
in Fig. 8a.

In non-resonant junctions, the currents $\tilde K_{-n}$ in Eq. (\ref{In})
have the singularities which are responsible for the current onset at
$eV=2\Delta$ and subharmonic gap structure. In resonant junctions, the
singularities are washed out due to strong electron-hole dephasing and
the resonant transmissivity is simultaneously renormalized.
In the case of a single-particle current in Eq. (\ref{I1}), the onset of 
non-resonant current is caused by zeros of the function $P_1$.
Calculation of $P_1$ for the resonant junctions by using the rule in 
Eq. (\ref{P'}) and retaining only
the resonant scattering amplitude $d^-_0$ yields
\begin{equation}
{D_0^-\over P_1}\approx {\Delta^4\Gamma^2\over 16\xi^2\xi_{-1}^2} \left|
E_0^- +{i\over 2}
\left(
\Gamma_{0}+\Gamma_{-1}
\right)\right|^{-2},
\label{P1}
\end{equation}
where
$\Gamma_{n}=\Gamma |E_{n}|/\xi_{n}$.
Equation (\ref{P1}) shows the transformation of the resonant tunneling
probability in the superconducting junctions: the resonance width is
broadened due to superconducting density of states $E/\xi$. Taking into
account Eq. (\ref{P1}) and similar equations for the other terms in Eq.
(\ref{I1}), we present the single-particle current on the form of the
Landauer formula,
\begin{equation}
I_1={e\over\pi}
\int_{\Delta}^{eV-\Delta} dE\,\tilde D_1(E), \label{I1'}
\end{equation}
with the effective single-particle transmission coefficient,
\begin{equation}
\tilde D_1(E)=
{\Gamma_0\Gamma_{-1}\over|E_0^- +(i/2)(\Gamma_{0}+\Gamma_{-1})|^2}. 
\label{D1}
\end{equation}
Similar equation have been derived in Ref. \cite{Ye2} using a different method.

Equations (\ref{I1'}) and (\ref{D1}) determine the current in the wedge
region in Fig. 6a. In the limit of $\Gamma\rightarrow 0$, the resonant
current reads
\begin{equation}
I_1={2e^2\Gamma
V_+ V_-\theta[eV-2(\Delta+|E_0|)]
\over V_-\sqrt{(eV_+)^2-4\Delta^2}+V_+\sqrt{(eV_-)^2-4\Delta^2}} , 
\label{I1res}
\end{equation}
where $eV_{\pm}=eV\pm 2|E_0|$. This equation quantitatively describes the
single-particle current feature in Fig. 3. The current has maximum at
the wedge edges
and decreases at large $eV$ approaching the value for the resonant current
in the normal junction $I_N=e\Gamma$ (see Fig. 3),
\begin{equation}
I_1=I_N
\left\{
\begin{array}{rl}
\displaystyle
{2|E_0|+\Delta\over\sqrt{|E_0|(|E_0|+\Delta)}}, & eV=2(\Delta+|E_0|)\\
\displaystyle
1+{2\Delta^2\over (eV)^2} , & eV\gg \Delta, E_0 \end{array}
\right.
\label{I1asym}
\end{equation}
The current peak is the result of enhancement of the effective width of the
resonance in Eq. (\ref{D1}) at low energy $\xi=0$
Equation (\ref{I1asym}) is everywhere applicable except of the wedge vertex,
$E_0=0$, $eV=2\Delta$, where the current grows without limit. In fact, the
current consists of the peak and turns to zero at $eV=2\Delta$ due to
shrinking of interval of integration in Eq. (\ref{I1'}) in this region. The
maximum current is achieved when the integration interval becomes
comparable with the resonance width, $eV-2\Delta\sim
\Gamma\sqrt{\Delta/(eV-2\Delta)}$. These arguments yield estimate for the
maximum current at $eV=2\Delta$,
$(I_1)_{max}\sim I_N(\Delta/\Gamma)^{1/3}$.

\subsection{Pair current}
The pair current has form
\begin{eqnarray}
I_2={4e\over \pi}
\int_{\Delta}^{2eV-\Delta} dE\,
{|E_{-2}|\xi\xi_{-2}\over \Delta^3} \nonumber \\
\left\{
D_0^-D_{-2}^+
\left(
{e^{-\gamma}\varphi_{-2}\over P_2}
+{e^{\gamma}\bar\varphi_{-2}\over \bar P_2} \right)+
\left(
E_0\rightarrow -E_0
\right)
\right\}.
\label{I2}
\end{eqnarray}
Restricting again the consideration with quasiparticles incoming from the
left, we find that 
this current is contributed by two resonances, $E_0^-=0$ and $E_{-2}^+=0$,
which simultaneously enter the integration interval within the region
$II_1$ in Fig. 7 (region $II_2$ corresponds to resonant quasiparticles
incoming from the right).
Therefore, the resonant pair current only appears if the normal resonance
is sufficiently far from the Fermi level, $E_0>\Delta/2$, while at
$E_0<\Delta/2$ the current is non-resonant within the voltage interval
$\Delta<eV<2\Delta$. This means in particular that the onset of the pair
current at $eV=\Delta$ is small: $I_2\sim I_N(\Gamma/\Delta)^3$ if $E_0=0$.
In regions $II$, the pair current undergoes resonant enhancement, $I_2\sim
I_N(\Gamma/\Delta)^2$ due to independent contributions of two separate
resonances (Fig. 8b), each contribution being described by the equations 
similar to Eqs. (\ref{I1'}), (\ref{D1}).

The most interesting phenomenon in the resonant pair current is the overlap
of the resonances occurring along the lines $eV=\pm 2|E_0|$ in Fig. 7.
The overlap of the resonances produces a huge
current peak near these lines (seen as light streaks in the Fig 6a; we note
that these lines correspond to the position of the onset of resonant
current in normal junctions). The
scattering diagram for this case is presented in Fig. 8c. Applying Eq.
(\ref{P'}) for calculation of $P_2$ and retaining both the resonant
amplitudes $d_0^-$ and $d_{-2}^+$, we obtain
\begin{eqnarray}
{D_0^-D_{-2}^+\over P_2} \approx
{\Delta^6\Gamma^4\over |8\xi\xi_{-1}\xi_{-2}|^2} \left|
\left[E_0^- +i\left({\Gamma_{0}+\Gamma_{-1}\over 2}\right)\right]
\right. \nonumber\\
\left. \left[E_{-2}^+ +i\left({\Gamma_{-1}+\Gamma_{-2}\over 2}\right)\right]
-{\Gamma^2\Delta^2\over 4|\xi{-1}|^2}
\right|^{-2}.
\label{P2res}
\end{eqnarray}
Substituting Eq. (\ref{P2res}) into Eq. (\ref{I2}) for the current and
collecting the contributions of all scattering modes, we find %
\begin{equation}
I_2={e\over \pi}
\int_{\Delta}^{2eV-\Delta} dE\,
\tilde D_2(E),
\label{I2'}
\end{equation}
where
\begin{eqnarray}
&&\tilde D_2(E)= \nonumber \\
&&{\Gamma_{0}\Gamma_{-2}\Gamma^2\Delta^2/4|\xi_{-1}|^2\over
|\tilde E_{0}^-\tilde E_{-2}^+ -(\Gamma_{0}\Gamma_{-2}
+\Gamma^2\Delta^2/|\xi_{-1}|^2)/4+
i(\Gamma_{-2}E_{0}^- + \Gamma_{0}E_{-2}^+)/2|^2} \nonumber \\
&&\tilde E=
E+i\Gamma_{-1}/2.
\label{D2}
\end{eqnarray}
Equation (\ref{D2}) shows a remarkable similarity to the resonant
transmissivity of Schr\"odinger three-barrier structures: the probability to
leak outside the superconducting gap
through the sidebands $n=0$ and $n=-2$ (Fig. 8c) corresponding to
probability of tunneling through side barriers, while the probability of
Andreev reflection by the sideband $n=-1$ corresponding to transmissivity 
of a central barrier. Such three-barrier structures have been investigated, 
e.g. in connection with normal electron transport properties of
coupled quantum dots \cite{Naz,Delft}. The strong overlap of the resonances
is explained by the fact that shift of the resonances is proportional to 
$\Gamma^2$ due to Andreev reflection, according to Eq. (\ref{D2}), while the 
resonance width is proportional to the first power of $\Gamma$ (the quantity
$\Gamma_{-1}$ is equal to zero at the lines $eV=\pm2|E_0|$). 

In the vicinity of the lines $eV=\pm 2|E_0|$ and in the limit
$\Gamma\rightarrow 0$, the pair current has following form %
\begin{equation}
I_2=I_N{\Gamma^2 eV\sqrt{(eV)^2-\Delta^2}\over (eV-2|E_0|)^2
[(eV)^2-\Delta^2]+\Gamma^2[2(eV)^2-\Delta^2]}.
\label{I2double}
\end{equation}
(We notice that this formula is valid at all voltages $eV>\Delta$ because the 
side band
$n=-1$ is inside the energy gap if $eV\approx \pm 2|E_0|$). Equation
(\ref{I2double}) describes the current peak in Fig. 3, the height of the peak
\begin{equation}
(I_2)_{max}=I_N\;{2|E_0|\sqrt{4E_0^2-\Delta^2}\over 8E_0^2-\Delta^2}
\label{I2p}
\end{equation}
being comparable to the magnitude of the resonant single-particle current,
in particular, $(I_2)_{max}=I_N/2$ at $E_0\gg\Delta$.

According to Eq. (\ref{I2double}) the resonant pair current tends to zero
at large voltage $eV\gg\Delta, E_0$, which means that, rigorously speaking,
there is no resonant excess current. However, if the resonance is far
beyond the gap, $|E_0|\gg\Delta$, the current may strongly deviate from the
current in the normal junction in the region $\Delta\ll eV\ll 2|E_0|$
because the single-particle current is non-resonant in this region, while
the pair current is resonant. Such an effect is particularly pronounced in
the junctions where the resonance level follows the chemical potential of
one of the electrodes, $E_0(eV)\pm eV/2 \approx \epsilon=const$. The IVC in
this case corresponds to cuts in the plot in Fig 6a parallel to the light
streaks. In such a case, the peak of the pair current is very broad, and
even transforms into a plateau with a sharp onset at $eV=\Delta$
($\epsilon=0$), as shown in the inset in Fig. 10. The magnitude of the
current at the plateau can be found directly from Eq. (\ref{Kexp}) when
assuming $E_0=\epsilon\pm eV/2$ and $eV=\infty$, %
\begin{eqnarray}
I_2(\epsilon, \Gamma)={2e\over \pi}
\int_{0}^{\infty}dE\;\cosh(Re\gamma) \nonumber \\
{2D_0^- \sinh(Re\gamma) + D_0^-D_0^+ e^{-Re\gamma}\over
|e^{\gamma}-r_0^{-\ast}r_0^+e^{-\gamma}|^2}, 
\end{eqnarray}
$D_0^{\pm}=\Gamma^2/[((E\mp\epsilon)^2+\Gamma^2]$. This current as function
of $\epsilon$ is shown in Fig. 10.

There is an interesting difference between the property of the resonance in
the single-particle current and that of individual resonances of the pair
current. To be specific, let us consider the resonance $E^-_0$: in the pair
current this resonance is more narrow because the quantity $\Gamma_{-1}$ is
imaginary and causes a resonant shift
rather than a contribution to the resonance width. The physical reason for
this squeezing of the resonance is that direct leakage of a quasiparticle
through the side band $n=-1$ is blocked, and the only escape from the
resonant region into the continuum is through the states of the side band
$n=0$.

\subsection{High-order currents}

The effect of the resonance narrowing is even more important
for the third order current,
\begin{eqnarray}
&&I_3={4e\Delta\over \pi}
\int_{\Delta}^{3eV-\Delta} dE\,
{|E_{-3}|\xi\xi_{-3}\over \Delta^3} \nonumber \\
&&\left\{
D_0^-D_{-2}^+D_{-2}^-
\left(
{e^{-\gamma}\varphi_{-3} \over P_3}+
{e^{\gamma}\bar\varphi_{-3}\over \bar P_3} \right)+\left(
E_0\rightarrow -E_0 \right) \right\}.
\label{I3}
\end{eqnarray}
The third order current  has three resonances at $E_0^-, E_{-2}^{+},
E_{-2}^{-}=0$ which belong to the interval of integration within the
regions $III_1, III_2, III_3$ in Fig. 7, respectively. The side resonances
at $E_0^-, E_{-2}^{-}=0$ are characterized by an effective transmissivity
similar to the effective transmissivity of the resonances of pair current
(times additional factor $\sim \Gamma ^2$). The contribution of these
resonances in the current is therefore estimated as $I_3\sim
I_N(\Gamma/\Delta)^4$. The central resonance $E_{-2}^{+}=0$ is much more
narrow. Indeed, in this case (Fig. 8d), direct leakage of the resonant
particle into continuum is blocked at the both sidebands $n=-1,-2$, and the
particle can escape only through the sideband states $n=0,-3$, traversing
the junction one more time. The central resonance determines the current in
the vicinity of the threshold $eV=2\Delta/3,\, E_0=0$.

Calculation of the quantity $P_3$ in region $III_2$ according to Eq.
(\ref{P'}) yields
\begin{equation}
{D_0^-D_{-2}^+D_{-2}^-\over P_3} \approx
{\Delta^4\tilde\Gamma_0\tilde\Gamma_{-3}\over |4^2\xi\xi_{-3}EE_{-3}|}
\left|
\tilde E_{-2}^+ +{i\over2}
\left(\tilde\Gamma_{0}+
\tilde\Gamma_{-3}
\right)\right|^{-2}
\label{P3res}
\end{equation}
where $\tilde E_{-2}^+=E_{-2}^+ +i(\Gamma_{-1}+\Gamma_{-2})/2 +
O(\Gamma^2)$ and $\tilde\Gamma_{0}=\Gamma_{0}D_{0}^-\Delta^2/4|\xi|^2$,
$\tilde\Gamma_{-3}=\Gamma_{-3}D_{-2}^-\Delta^2/4|\xi_{-2}|^2$. According to
Eq. (\ref{P3res}), the resonance width is of the order of
$\tilde\Gamma\sim\Gamma^3$ which yields giant enhancement of the current
$I_3\sim I_N(\Gamma/\Delta)^2$, exceeding by two orders of $\Gamma$ the 
contribution of the side resonances. Such narrowing
of the central resonance occurs in the quadrangle region in Fig. 7 bounded
by the edges of the resonance region $III_2$ and regions $II$. The current
in this region
has a form similar to the one in equation (\ref{I1'}),
\begin{equation}
I_3={e\over \pi}
\int_{\Delta}^{3eV-\Delta} dE\,
\tilde D_3(E),
\label{I3'}
\end{equation}
with the effective resonant transmissivity %
\begin{equation}
\tilde D_3(E)={3\tilde\Gamma_0\tilde\Gamma_{-3}
\over
\left|
\tilde E_{-2}^+ - i
\left(
\tilde\Gamma_{0}+\tilde\Gamma_{-3}
\right)/2
\right|^2}.
\label{D3}
\end{equation}
In the limit of small $\Gamma\rightarrow 0$, the current reads %
\begin{equation}
I_3=6e
\tilde\Gamma_{0}
\tilde\Gamma_{-3}/
\left(
\tilde\Gamma_{0}+\tilde\Gamma_{-3}\right)_{E=|E_0|+3eV/2}. \label{I3r}
\end{equation}

The phenomenon of resonance narrowing
provides the explanation for the absence of current structure at voltage
$eV=\Delta$, namely the dominance of the third order current $I_3$ at the
threshold of the pair current. The current in Eq. (\ref{I3r}) is
responsible for the light wedge-like region at $eV<\Delta$ in Fig. 6b.
Similarly to the case of single-particle current, the third-order current
in Eq. (\ref{I3r}) has a peak at the edges of the wedge with the height
increasing proportionally to $(eV-2\Delta/3)^{-1/2}$ towards the vertex of
the wedge, $eV=2\Delta/3, \, E_0=0$. This growth is again limited due to
interplay between shrinking integration interval and growing resonance
width,
$eV-2\Delta/3\sim \Gamma^3[\Delta/(eV-2\Delta/3)]^{1/2}$. This estimate
gives a height $(I_3)_{max}\sim I_N(\Gamma/\Delta)$ of the current peak at
$eV=2\Delta/3$.
As one may see in Fig. 6b, there are no current structures at the edges
$eV=2(\Delta-|E_0|)$ of the above mentioned quadrangle where the the
narrow resonance of three-particle current dies: this is because of the 
resonant pair current
emerges at the same lines, giving rise to a gradual cross over between 
three-particle current and pair current both having the magnitude of the
order of $I_N(\Gamma/\Delta)^2$.

The phenomenon of resonance narrowing results in enhancement of central
resonances in all higher odd-order currents, giving rise to the current
peaks at $eV=2\Delta/(2k+1)$, $E_0=0$ with the height $I_{max}\sim
I_N(\Gamma/\Delta)^{2k-1}$. The magnitude of the current between
neighbouring peaks is $I\sim I_N(\Gamma/\Delta)^{2k}$. Also, the overlap of
narrow resonances of the even-order currents near the lines $eV=\pm
2|E_0|$ yields
current peaks with the height $I\sim I_N(\Gamma/\Delta)^{2k}$ within the
intervals
$2\Delta/(2k+2)<eV<2\Delta/2k$. These current peaks are clearly seen in
Fig. 6b in the form of light streaks.

\section{Conclusion.}

In conclusion, we have considered effect of the normal electron resonant
tunneling on the subharmonic gap structure (SGS) in  mesoscopic
superconducting junctions. In non-resonant tunnel junctions, the SGS
consists of sharp onsets and narrow peaks of the current at voltage
$eV=2\Delta/n$. In resonant junctions, SGS is considerably modified
depending on the position of resonance level with respect to the chemical
potentials of the electrodes. If the resonance level is situated exactly
in the middle between the chemical potentials of electrodes, the odd-$n$
current structures are tremendously enhanced while the even-$n$ current
structures are not affected by the resonance. This enhancement is explained
by narrowing of the resonance during multiple Andreev reflections. When the
resonance departs from the midpoint between the chemical
potentials of electrodes, new current structures appear at $eV=\pm2E_0$ in
a form of current peaks. This feature results from overlap of electron and
hole resonances. 

In our calculations, the Coulomb charging energy was assumed to be smaller
than the superconducting gap, and the charging effects were neglected. In
experiments on  metallic dots \cite{Ralph} and carbon nanotubes 
\cite{Delft1,Rice,Dekker}, the opposite situation has been observed with 
the Coulomb charging energy exceeding the superconducting gap which led
to suppression of the subgap current. The charging energy in quantum
transport experiments can
be reduced by enhancing capacitance of the resonant structure, e.g. by using
substrates with large dielectric constants. This will allow direct
application of our results  to such structures. Another way would be to use
high-$T_c$ materials for fabrication of superconducting electrodes for
the nanotube experiments. Our theory is applicable to ballistic plane
junctions with large capacitance such as resonant junctions in high
mobility S-2DEG-S devices and atomic plane junctions in layered cuprates
(intrinsic Josephson junctions \cite{Kleiner}). Current-voltage
characteristics of such multimode junctions can be obtained on the basis of
our theory by summation of contributions of all transport modes.

\section{Acknowledgement}

This work has been supported by the Swedish Natural Science Research
Council (NFR), the Swedish Board for Technical Development (NUTEK), the
Swedish Royal Academy of Sciences (KVA), and the New Energy Development
Organization (NEDO), Japan.


\appendix
\section{Derivation of current}

Following the method of Ref. \cite{Sh1,Sh2}, we eliminate the Andreev
scattering amplitudes $a_n$ and $b_n$ from Eqs. (\ref{mar}) for the
scattering state $j=1$ and obtain a closed set of equations for the normal
amplitudes $c_n$ and $f_n$,
\begin{equation}
c_{2n} +V^-_{2n+1} f_{2n+1} + V^+_{2n-1} f_{2n-1} ={2\xi\over \Delta
\Xi_0}\delta_{n0}
\label{cf}
\end{equation}
$$
f_{2n+1} +V^+_{2n+2} c_{2n+2} + V^-_{2n} c_{2n} =0. $$
The coefficients in these equations are
\begin{eqnarray}
V^-_{2n}&=& d_{2n}^{+\ast}e^{(\gamma_{2n}+\gamma_{2n+1})/2}/\Xi_{2n+1},
\nonumber \\
V^+_{2n}&=&-\sigma_{2n}\sigma_{2n-1}
d_{2n}^{-\ast}e^{-(\gamma_{2n}+\gamma_{2n-1})/2}/\Xi_{2n-1},
\nonumber \\
V^-_{2n+1}&=&d_{2n}^{+\ast}e^{(\gamma_{2n}+\gamma_{2n+1})/2}/\Xi_{2n},
\nonumber \\
V^+_{2n-1}&=&-\sigma_{2n}\sigma_{2n-1}
d_{2n}^{-\ast}e^{-(\gamma_{2n}+\gamma_{2n-1})/2}/\Xi_{2n} \nonumber ,
\end{eqnarray}
where the quantities $\Xi_n$ are defined as
\begin{eqnarray}
\Xi_{2n}&=&r_{2n}^{+\ast}e^{\gamma_{2n}}-r_{2n}^{-\ast}e^{-\gamma_{2n}},
\nonumber \\
\Xi_{2n-1}&=&r_{2n-2}^{+\ast}e^{\gamma_{2n-1}}-r_{2n}^{-\ast}e^{-\gamma_{2n-1}}. 
\label{Xi}
\end{eqnarray}
In non-resonant junctions, the functions $\Xi_n$ approach $r\xi_n$ since
energy dispersion of the reflection amplitude is negligibly small.
SGS in non-resonant junctions is caused by zeros of the
functions $\xi_n$ \cite{Sh2}, and renormalization of these functions in the
resonance case, $\xi_n\rightarrow\Xi_n$, is the reason of considerable
difference of the SGS in resonant junction compared to the SGS in non-resonant
junctions.
We solve equations
(\ref{cf}) by introducing ratios $S_{2n}=c_{2n}/f_{2n-1}$ and $S_{2n+1}=
f_{2n+1}/c_{2n}$ and expressing $f_n$ through $c_0$, %
\begin{equation}
f_n=\prod_{i=0}^n S_ic_0,
\label{fn}
\end{equation}
$c_0$ being related to $S_{\pm 1}$ by virtue of the first equation in Eq.
(\ref{cf}). By introducing chain-fractions $Z_n$, \begin{equation}
Z_{0}=1-(d_0^{+\ast})^2{e^{\gamma+\gamma_1}\over \Xi_0 \Xi_{1}Z_1}-
(d_0^{-\ast})^2{e^{-\gamma-\gamma_{-1}}\over \Xi_0 \Xi_{-1}Z_{-1} },
\label{Z}
\end{equation}
\begin{eqnarray}
Z_{2n}&=&1- (d_{2n}^{-\ast})^2{e^{-\gamma_{2n}-\gamma_{2n-1}}\over
\Xi_{2n}\Xi_{2n-1}Z_{2n-1}}, \nonumber \\
Z_{2n-1}&=&1- (d_{2n-2}^{+\ast})^2{e^{\gamma_{2n-1}+\gamma_{2n-2}}\over
\Xi_{2n-1}\Xi_{2n-2}Z_{2n-2}} \nonumber  ,
\end{eqnarray}

one can rewrite Eq. (\ref{fn})
for $f_n$, e.g. with negative sideband index $n=-2k-1<0$, in the following form
\begin{equation}
\hspace{-1cm}
f_{-2k-1}=
{2\xi\over\Delta }(-1)^k
d_{0}^{-\ast}e^{-(\gamma_{}+\gamma_{-2k-1})/2} \prod_0^{2k+1}{\sigma
_{-i}\over \Xi_{-i}Z_{-i}} \prod_1^{k} d_{-2i}^{+\ast}d_{-2i}^{-\ast}.
\label{f}
\end{equation}
An equation for $b_{-2k-1}$ follows from
Eqs. (\ref{mar}) and (\ref{f}),
\begin{equation}
\hspace{-1cm}b_{-2k-1}= r_{-2k-2}^{+\ast}e^{\gamma_{-2k-1}} \left(
1+{D_{-2k-2}^{+}e^{\gamma_{-2k-2}}\over
r_{-2k-2}^{+}\Xi_{-2k-2}Z_{-2k-2}}\right)f_{-2k-1}, \label{b}
\end{equation}
where $D_n^\pm=|d_n^\pm|^2$.

Collecting the normal and Andreev transmission amplitudes in Eqs. (\ref{f})
and (\ref{b}) and substituting them into Eq. (\ref{I}) for the current, we
finally get
\begin{eqnarray}
&&K_{n}=-|f_{n}|^2\cosh(Re\gamma_{n}) \nonumber \\
&&\left( 1- e^{2Re\gamma_{n}}(1-D_{n-1}^{+})
\left|
1+{D_{n-1}^{+}e^{\gamma_{n-1}}\over
r_{n-1}^{+}\Xi_{n-1}Z_{n-1}}\right|^2
\right),\ n<0.  \nonumber \\
\label{Kexp}
\end{eqnarray}
The corresponding equation for positive $n>0$ is obtained from Eq.
(\ref{Kexp}) via the substitutions $\gamma\rightarrow -\gamma$,
$D^+\rightarrow D^-$, and $n-1\rightarrow n+1$.
%
\section{Transformation of $K_n$}
Equation (\ref{Kexp}) is not convenient for analysis of the subharmonic gap
structure \cite{Sh2}. The current structures are caused by the processes of
creation of real excitations during across-the-gap transitions
$E\rightarrow E_n<-\Delta$. The currents of other side-bands ($E_n>\Delta$)
perfectly cancel each other at zero temperature \cite{BB}. Furthermore,
there is rigorous balance between Andreev and normal channel currents among
the states lying within the superconducting gap \cite{BB}, which provides
successive drops of the current with decreasing applied voltage. All of
these features are not explicitly seen in the currents $K_n$
in Eq. (\ref{Kexp}). Moreover, the current structures related to the
creation of real excitations in even-order sidebands are hidden in Eq.
(\ref{K}) which consists of contributions from only odd
sidebands.

To overcome this difficulty, we will transform the sideband currents in Eq.
(\ref{Kexp}) following the method suggested in Ref. \cite{Sh2}. Assuming
the transparency $D_{n-1}^+$ in Eq. (\ref{Kexp}) to be small, we find that
the current $K_n$ to leading order 
is proportional to $\sinh(Re\gamma_{n})\sim \theta(E^2_{n}-\Delta^2)$.
This observation will allow us later [in Eq. (\ref{In})] to separate the
contributions from 
states below, $E_n<-\Delta$, and above, $E_n>-\Delta$ the energy gap.
Having separated out the leading term, we rewrite Eq. (\ref{Kexp}) in the
form

\begin{eqnarray}
&&K_{n}=|f_{n}|^2\cosh(Re\gamma_{n}) \nonumber \\
&&\left(
2\sinh(Re\gamma_{n})e^{Re\gamma_{n}}
-{D_{n-1}^{+}e^{2Re\gamma_{n}}\over
|\Xi_{n-1}Z_{n-1}|^2} F_{n-1}\right),
\end{eqnarray}
\begin{equation}
F_{n}= |\Xi_{n}Z_{n}|^2-
2Re\left(
e^{\gamma_{n}^\ast}r_{n}^{+}\Xi_{n}Z_{n} \right) - D_{n}^{+}e^{2Re\gamma_{n}}.
\label{Fn}
\end{equation}
Equation (\ref{Fn}) possesses a similar property: it is proportional to
$\theta(E^2_{n}-\Delta^2)$ to leading order with respect to $D$.
Separating out this leading term, we
further transform Eq. (\ref{Fn}) into the equation
\begin{equation}
F_{n}= -2\sinh(2Re\gamma_{n})-
{D_{n}^{-}e^{-2Re\gamma_{n}}\over
|\Xi_{n-1}Z_{n-1}|^2}G_{n-1},
\end{equation}
where
\begin{eqnarray}
G_{n-1} &=& |\Xi_{n-1}Z_{n-1}|^2 + \nonumber \\
&+& 2Re\left(
e^{-\gamma_{n-1}^\ast}r_{n}^{-}\Xi_{n-1}Z_{n-1} \right) -
D_{n}^{-}e^{-2Re\gamma_{n-1}}.
\end{eqnarray}
One more transformation,
\begin{equation}
G_{n-1}= 2\sinh(2Re\gamma_{n-1})-
{D_{n-2}^{+}e^{2Re\gamma_{n-1}}\over
|\Xi_{n-2}Z_{n-2}|^2}F_{n-2},
\end{equation}
accomplishes the cycle, yielding the quantity $F$ in Eq. (\ref{Fn}) with
shifted index.
Performing repeatedly such transformations, we get for the current in Eq.
(\ref{Kexp}) the following expansion
\begin{eqnarray}
\label{expKn}
\hspace{-.5cm}&K_{n}&=
\theta(E_{n}^2-\Delta^2)Q_{n} + \nonumber \\
&+& 2\theta(E_{n-1}^2-\Delta^2)e^{-Re\gamma_{n}}\cosh(Re\gamma_{n}) Q_{n-1}
+ \nonumber \\
&+& 2 \theta(E_{n-2}^2-\Delta^2)e^{-Re\gamma_{n}+2Re\gamma_{n-1}}
\cosh(Re\gamma_{n}) Q_{n-2} + \nonumber \\
&+& 2 \theta(E_{n-3}^2-\Delta^2)e^{-Re\gamma_{n}+2Re\gamma_{n-1}
-2Re\gamma_{n-2}}\cosh(Re\gamma_{n}) Q_{n-3}+ \nonumber \\
&+& ...
\end{eqnarray}

In this equation, the quantity $Q_{n}$ is defined as
\begin{eqnarray}
Q_{n}= {8\xi^2 \xi_{n}|E_{n}|e^{-\gamma} \over \Delta^4 P_n }
\,D_0^-\left(\prod_{i=1}^{k-1} D_{-2i}^+ D_{-2i}^- \right) D_{-2k}^+
\cdot \nonumber \\
\cdot \left \{
\begin{array}{cl}
1, & |n|=2k\\
D_{-2k}^-, & |n|=2k+1\\
\end{array}
\right.,
\label{Qn}
\end{eqnarray}
where
\begin{equation}
P_n=\prod_{i=0}^{|n|}|\Xi_{-i}Z_{-i}|^2. \label{P}
\end{equation}

Collecting together all terms with similar $\theta$-functions, we can
finally rearrange the sum in Eq. (\ref{I'}):
\begin{equation}
\sum_{n=odd}K_n\;\rightarrow\;\sum_{n=1}^\infty \tilde K_{n},\; \tilde
K_{n}=\theta(E_{n}^2-\Delta^2)Q_{n}\varphi_{n}, \label{Ktilde}
\end{equation}
where $\varphi_{n}$ is given by the recurrence equation $$
\varphi_{n-1}=1+ \exp[(-1)^{n+1} 2Re(\gamma_{n})]\varphi_{n},\;\;
\varphi_{-1}=1.
$$

Far from resonance, the quantity $P_n$ may cause strong singularity in the
sideband current due to the presence of zeros in the functions $\Xi_n$, and
accounting for the factors $Z_n$ is absolutely necessary for regularization
of the singularity \cite{Sh2}. In the resonant case, the functions $\Xi_n$
do not turn to zero because of strong
electron-hole dephasing, $r^+_n\neq r^-_n$, and the quantities $Z_n$ can be
omitted from Eq. (\ref{P}) in the limit of narrow resonance
$\Gamma\ll\Delta$,
\begin{equation}
P_n\approx{\prod_{i=0}^{n}}'|\Xi_{-i}|^2 \label{P'}
\end{equation}
The role of $Z_n$ in this limit reduces to cancellation of the terms in the
product (\ref{P'}) which are proportional to the squared resonance
amplitudes $(d_n^\pm)^2$; this is denoted by the prime in Eq. (\ref{P'}).
The presence of the resonant denominators in equation (\ref{Xi}) for
$\Xi$ gives rise to renormalization of the normal electron transmission
coefficients $D_n^\pm$ in Eq. (\ref{In}) for the current.

\begin{figure}
\psfig{figure=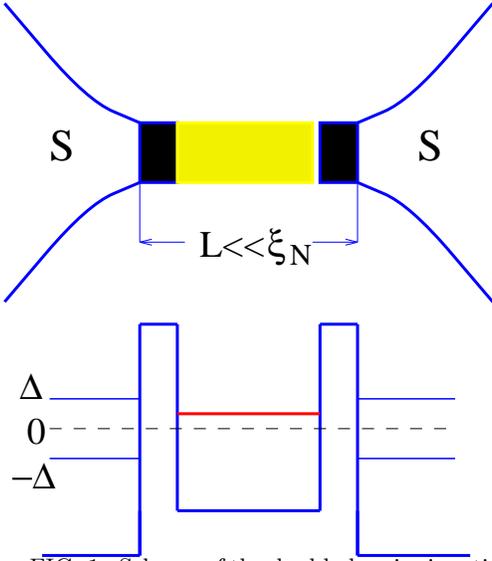,width=6.5cm}
\caption{Scheme of the double-barrier junction. Upper part: dark regions -
tunnel barriers; shadowed region - normal conductor. Lower part: energy
diagram showing one normal resonant level inside the superconducting gap.}
\end{figure}

\begin{figure}
\psfig{figure=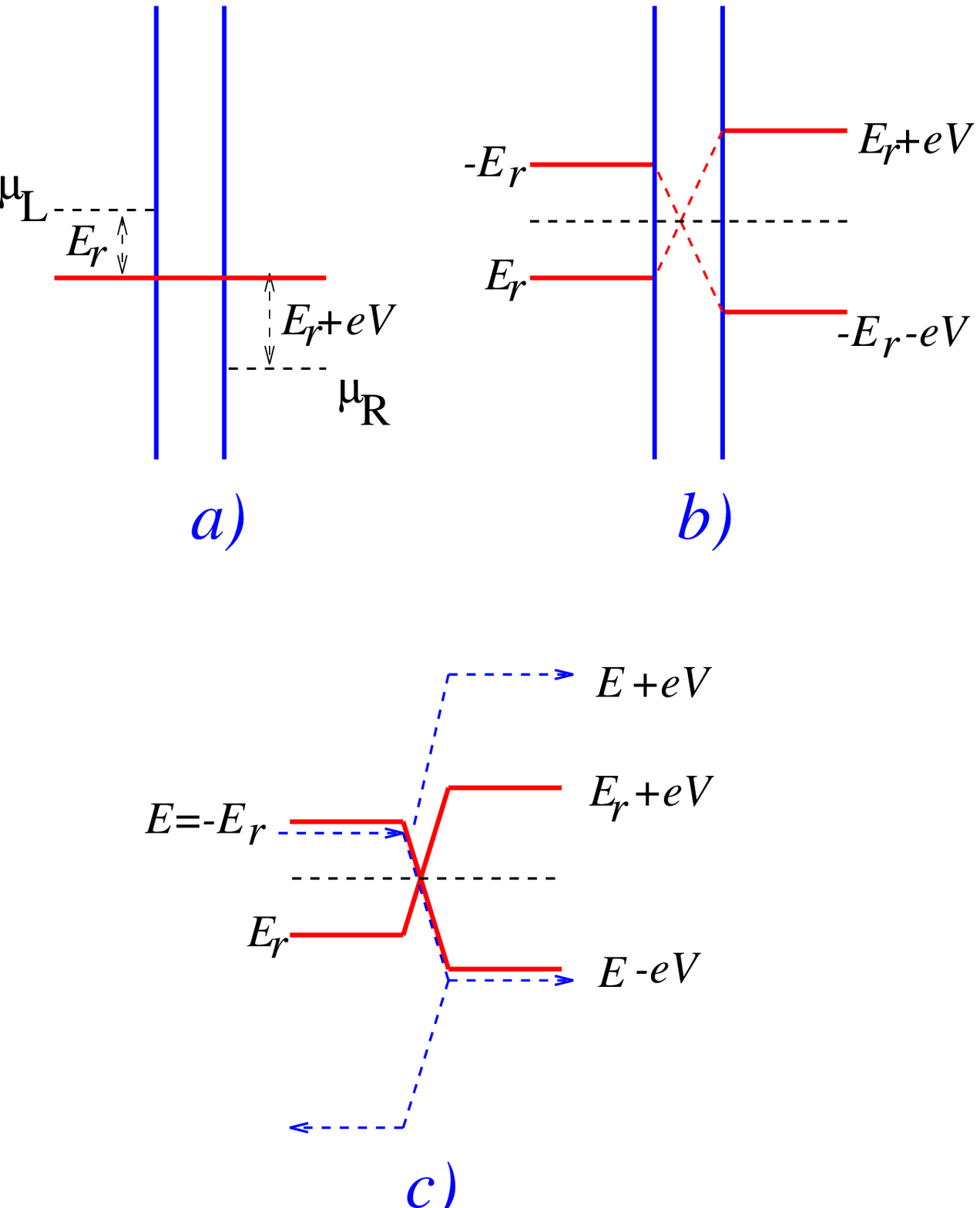,width=6.5cm}
\caption{Energy diagram of resonant junctions under applied voltage.
a) Resonance in normal junction; the distance to the chemical
potentials of the left and right electrodes is $E_r$ and $E_r+eV$
respectively.
b) Electron and hole resonances in superconducting junction; the
resonance offset is different in the left and right electrodes with 
respect to the global chemical potential after equalizing the chemical
potentials of the electrodes by means of the gauge transformation.
c) Resonant transition in the scattering diagram.}
\end{figure}

\begin{figure}
\psfig{figure=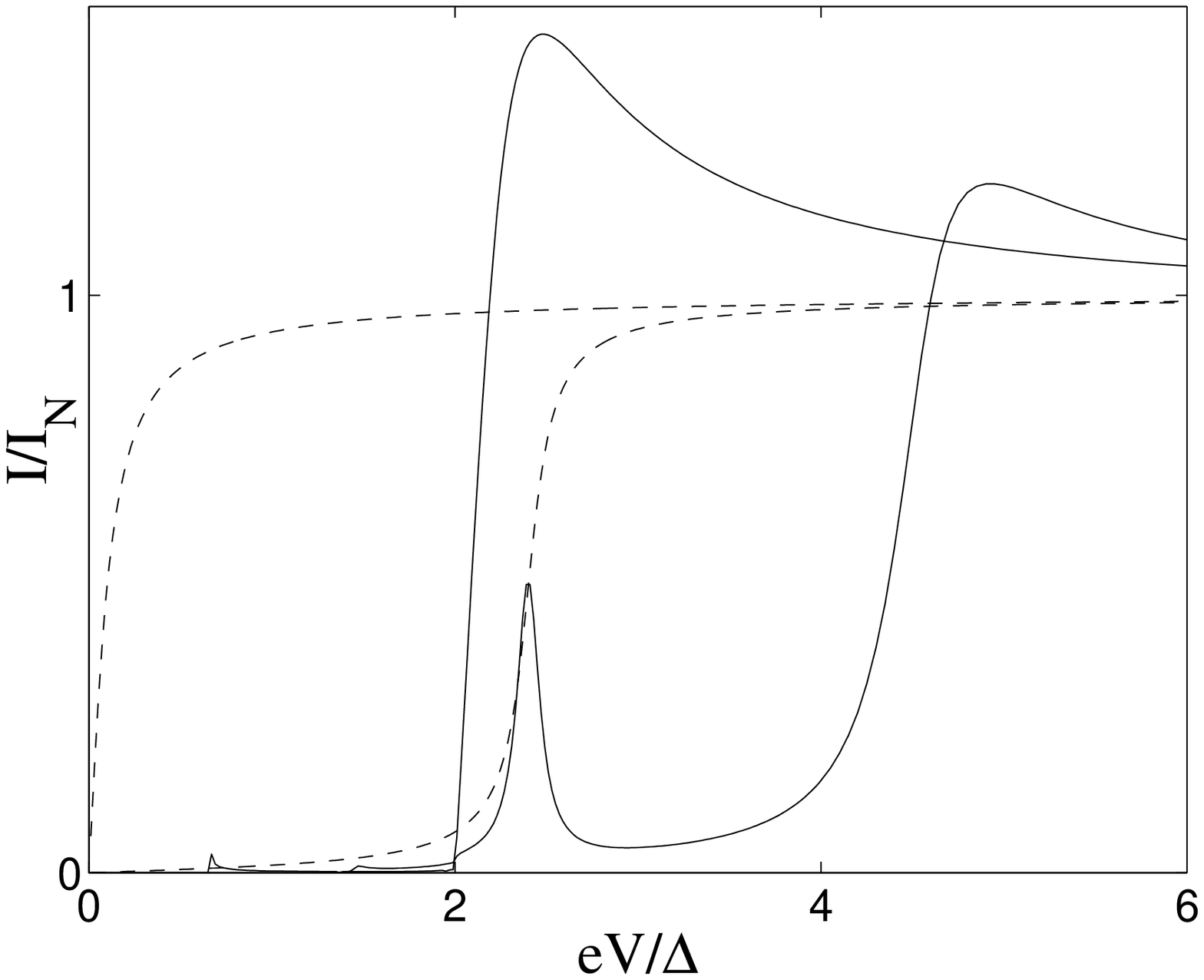,width=6.5cm}
\psfig{figure=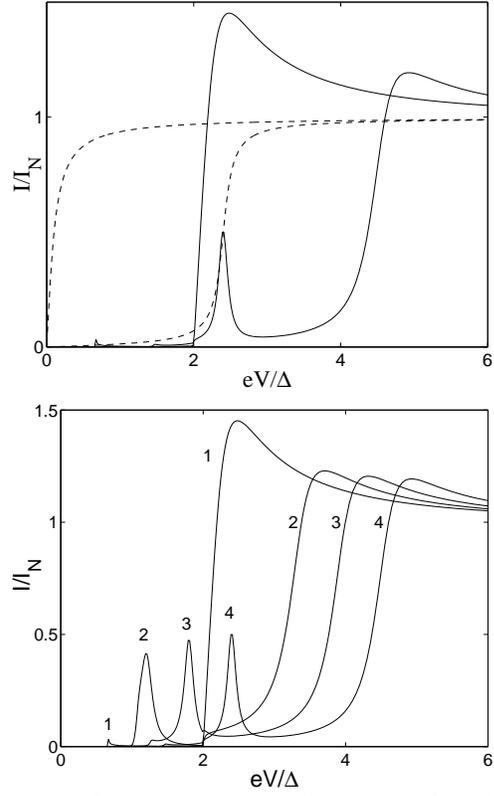,width=6.5cm}
\caption{IVC of symmetric resonant junctions.
a) IVC of normal junction (dashed line) and superconducting junction (solid
line), ${E_0}/{\Delta}$= 0, 1.2 (left and right curves respectively), 
$\Gamma/\Delta=0.05$. 
b) IVC of superconducting junction,  
${E_0}/{\Delta}=0.$ (1), 0.6 (2), 0.9 (3), 1.2 (4).}
\end{figure}

\begin{figure}
\psfig{figure=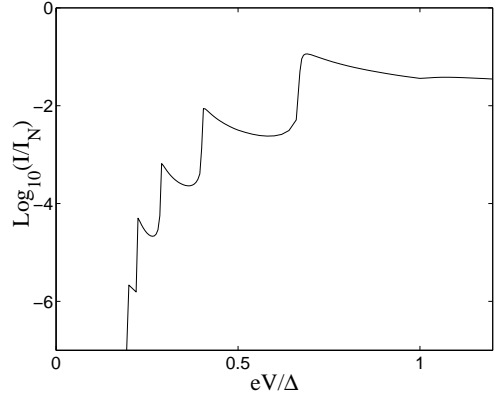,width=6.5cm}
\caption{Subharmonic gap structure on IVC of symmetric resonant junction, 
$E_{0}=0$,
$\Gamma/\Delta=0.2$. The structures appear only at $eV=2\Delta/n$ with odd
$n$.} 
\end{figure}

\begin{figure}
\psfig{figure=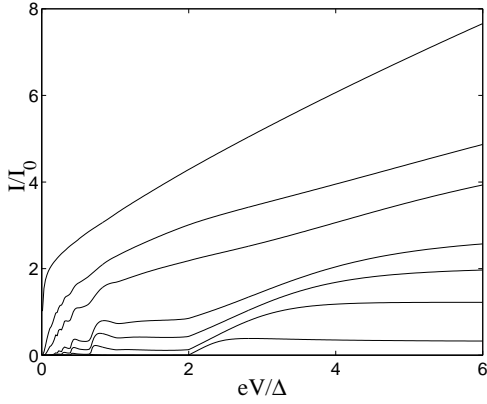,width=6.5cm}
\caption{ IVC of symmetric resonant junctions with $E_{0}=0$ and
${\Gamma}/{\Delta} \in \{0.1, 0.4, 0.7, 1.0, 2.0, 3.0, 10.0\}$ (from bottom
to top). The current is
normalized by $I_{0}=e\Delta/\hbar \pi$.
}
\end{figure}

\begin{figure}
\psfig{figure=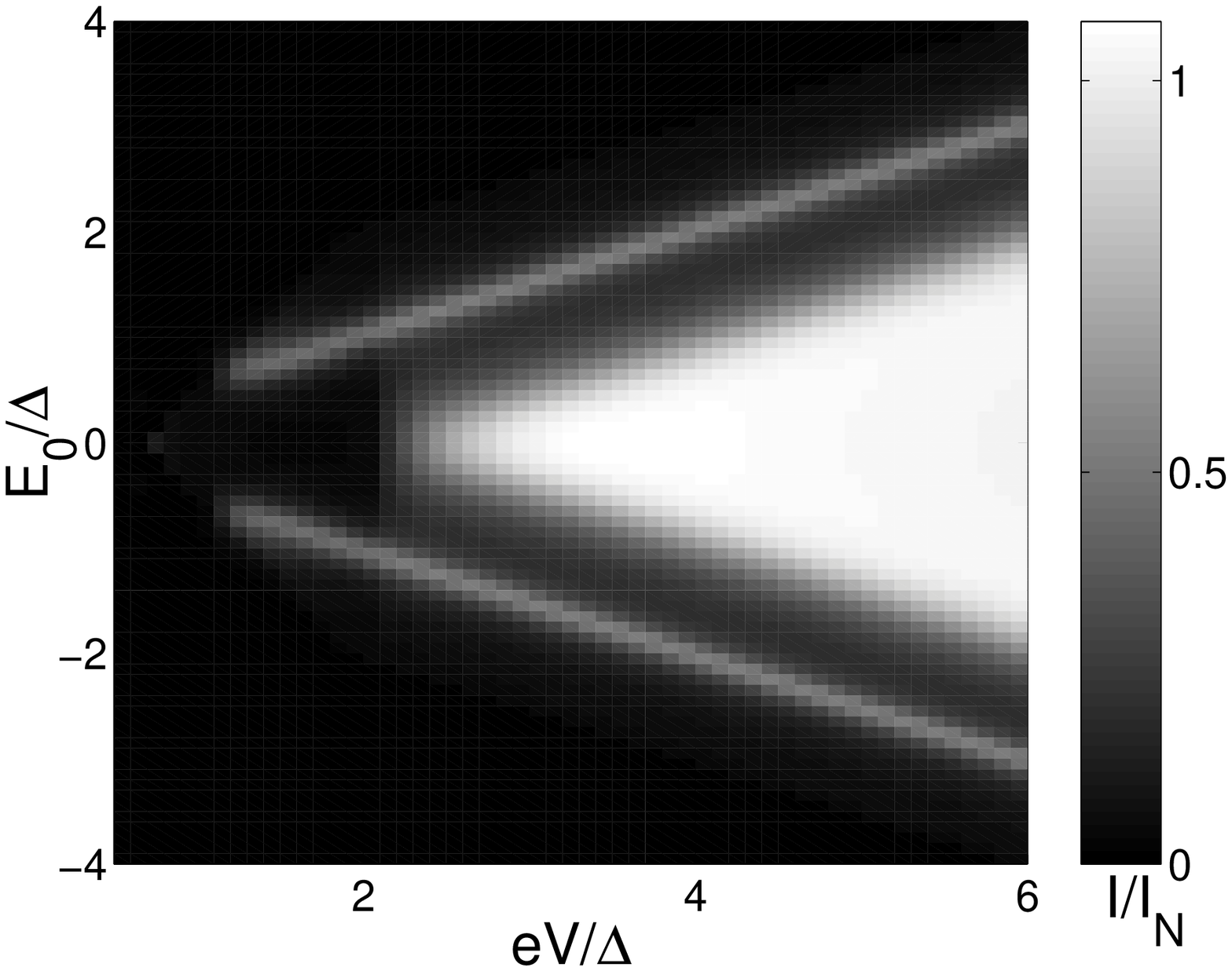,width=6.5cm }
\psfig{figure=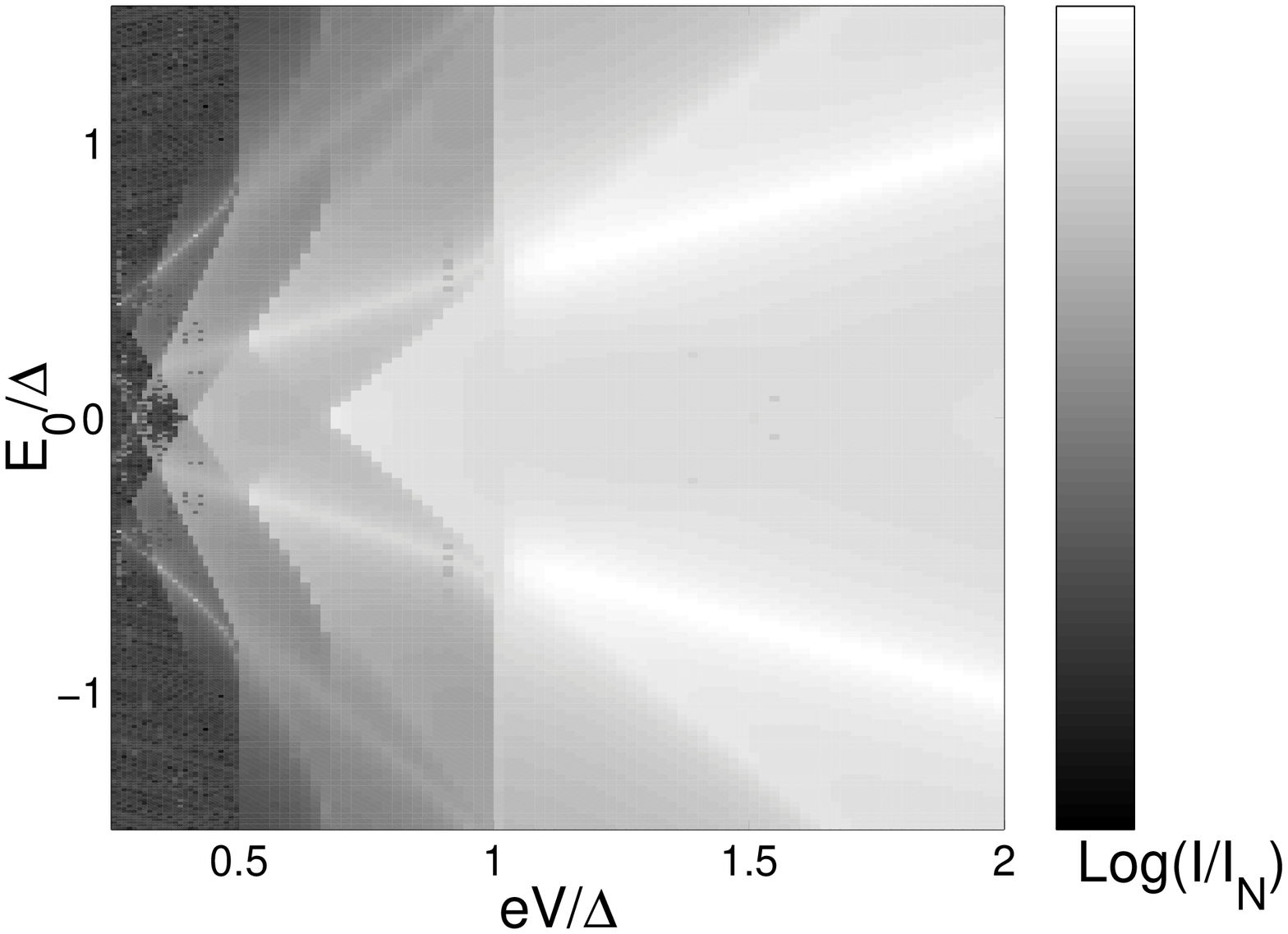,width=6.5cm}
\caption{Intensity plots showing the dependence of the current
$I(V, E_0) $ on the applied bias voltage $V$ and the resonance
position $ E_0$.
(a) Single-particle current and pair current at $eV>\Delta$
($\Gamma=0.2\Delta$).
(b) Pair current and high-order currents at $eV<2\Delta$
($\Gamma=0.05\Delta$).}
\end{figure}

\begin{figure}
\psfig{figure=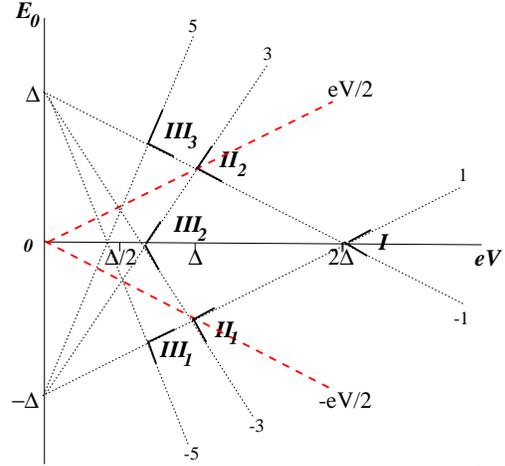,width=6.5cm}
\caption{. Resonant regions in the plane $(eV, E_0,)$ for sideband currents
$I_n$. $I$, $II$ and $III$ are  resonant regions for single-particle current,
pair current and three-particle current respectively. The resonant regions
are bound by dotted lines $E_0=\pm(\Delta-neV/2)$ (labeled with $\pm n$).
Bold dashed lines show positions of double resonances. }
\end{figure}

\begin{figure}
\psfig{figure=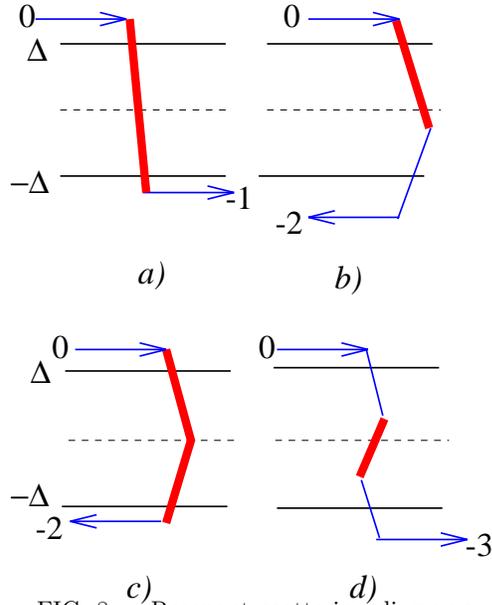,width=6.5cm}
\caption{ Resonant scattering diagrams; bold lines show resonant
transitions.
a) Resonance in single-particle current;
b) single resonance in the pair current; c) double resonance in the pair
current; d) central resonance in the three-particle current.} 
\end{figure}

\begin{figure}
\psfig{figure=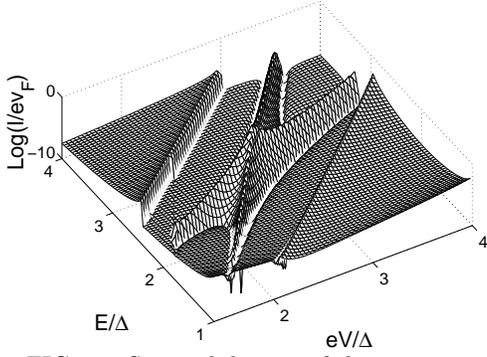,width=6.5cm}
\caption{ Spectral density of the pair current as a function of bias
voltage for $\Gamma/\Delta = 0.05$. The current peak appears at the crossing
point of two single resonances. }
\end{figure}

\begin{figure}
\psfig{figure=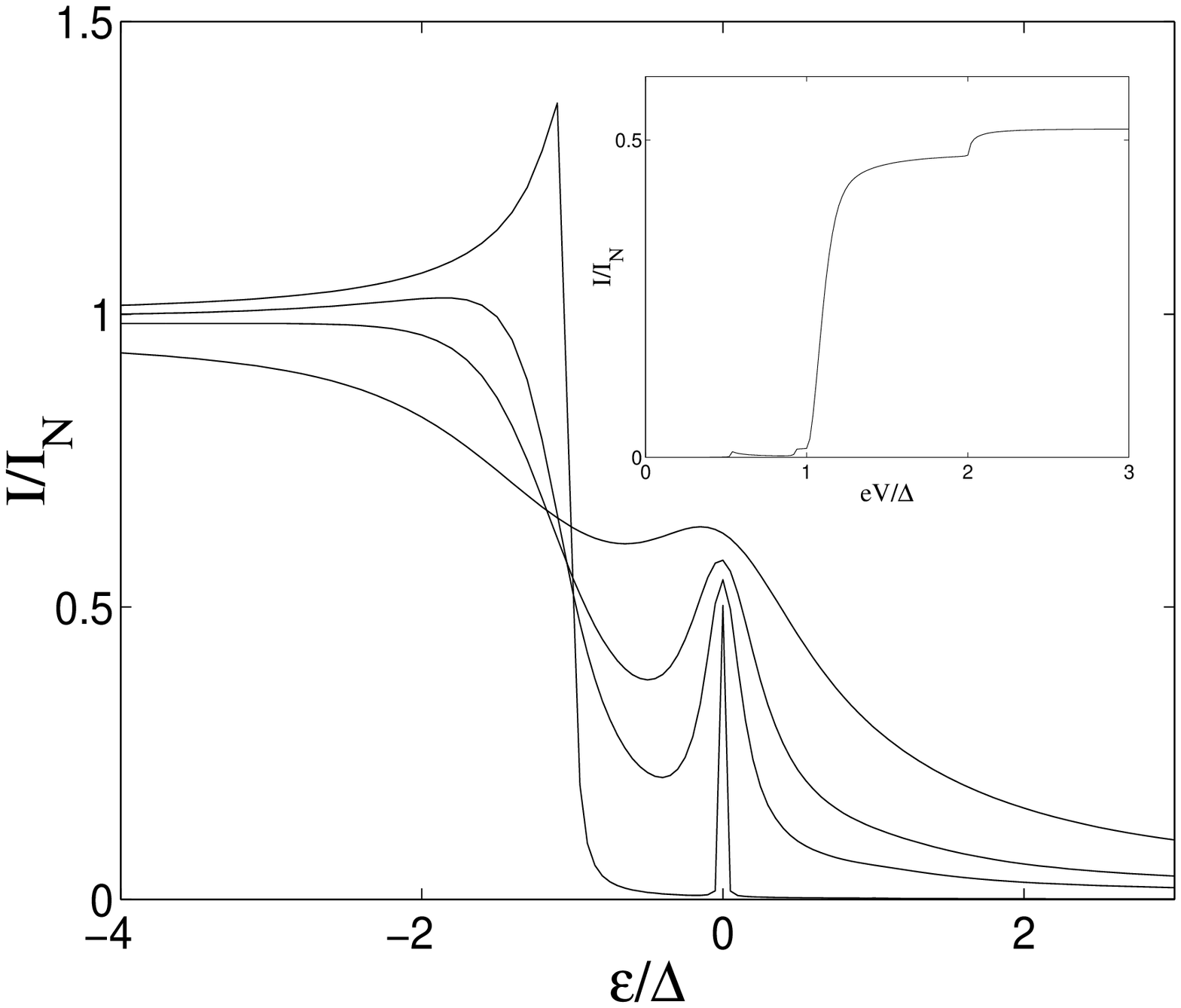,width=6.5cm}
\caption{Current for asymmetric junction at large voltage ($eV=1000\Delta$ 
as a function of the position
of the resonance $\epsilon = |E_0|-eV/2$ for different resonance width
$\Gamma/\Delta \in \{0.01, 0.21, 0.41, 0.61, 0.81, 1.01\}$ (from
top to bottom at the left side of the figure). Inset shows IVC for 
$\epsilon =0$: the resonance level coincides with the chemical potential of 
one of the electrodes.}
\end{figure}


\begin{thebibliography}{999}
\bibitem{Asl2}
L.G. Aslamazov and V.M. Fistul', Sov. Phys. JETP {\bf 56}, 655 (1982).
\bibitem{Tart}
A.T. Tartakovsky and V.M. Fistul', Sov. Phys. JETP {\bf 67}, 1695 (1988).
\bibitem{Gla}
L.I. Glazman and K.A. Matveev, JETP Lett. {\bf 49}, 659 (1989). 
\bibitem{Dev}
L.A. Devyatov and M. Yu. Kupriyanov, JETP Lett. {\bf 59}, 200 (1994).
\bibitem{Gla2}
I.L. Aleiner, Penny Clarke, and L.I. Glazman, Phys. Rev. B {\bf 53}, R7630
(1996).
\bibitem{Golub}
A. Golub, Phys. Rev. B {\bf 52}, 7458 (1995). 
\bibitem{Ovi}
A. Frydman and Z. Ovadyahu, Phys. Rev. B {\bf 55}, 9047 (1997).
\bibitem{Gross}
J. Harlbritter, Phys. Rev. B {\bf 46}, 11238 (1992). 
\bibitem{Asl}
L.G. Aslamazov and V.M. Fistul', Sov. Phys. JETP {\bf 54}, 206 (1981).
\bibitem{She}
I.F. Itskovich and R.I. Shekhter, Sov. J. Low Temp. Phys. {\bf 7}, 418 (1981).
\bibitem{Taka}
H. Takayanagi, T. Akazaki, and J. Nitta, Phys. Rev. Lett. {\bf 75}, 3533
(1995).
\bibitem{Yacobi}
A. Yacobi, M. Heiblum, D. Mahalu, and H. Shtrikman, Phys. Rev. Lett. {\bf
74}, 4047 (1995).
\bibitem{Ralph}
D.C. Ralph, C.T. Black, and M. Tinkham, Phys. Rev. Lett. {\bf 74}, 3241 (1995).
\bibitem{Delft1}
S.J. Tans et al., Nature {\bf 386}, 474 (1997).
\bibitem{Rice}
M. Bockrath et al., Science, {\bf 275}, 1922 (1997).
\bibitem{Volk}
P.I. Arseyev and B.A. Volkov, Solid State Comm. {\bf 78}, 373 (1991).
\bibitem{Bee}
C.W.J. Beenakker and H. van Houten, in {\em Single Electron Tunneling and
Mesoscopic Devices} (Springer, Berlin, 1991). \bibitem{Cre}
A. Chrestin, T. Matsuyama, and V. Merkt, Phys. Rev. B {\bf 49}, 498 (1994).
\bibitem{WSh}
G. Wendin and V.S. Shumeiko, Superlattices Microstruct. {\bf 20}, 569 (1996).
\bibitem{Sh1}
E.N. Bratus', V.S. Shumeiko, and G. Wendin, Phys. Rev. Lett. {\bf 74}, 2110
(1995).
\bibitem{Sh2}
V.S. Shumeiko, E.N. Bratus', and G. Wendin, Low Temp. Phys. {\bf 23}, 249
(1997).
\bibitem{Ave1}
D. Averin and A. Bardas, Phys. Rev. Lett. {\bf 75}, 1831 (1995).
 \bibitem{Ye1}
J.C. Cuevas, A. Martin-Rodero, and A. Levy Yeyati, Phys. Rev. B {\bf 54},
7366 (1996).
\bibitem{BB}
E.N. Bratus', V.S. Shumeiko, E.V. Bezuglyi, and G. Wendin, Phys.Rev. B {\bf
55}, 12666 (1997).
\bibitem{Jan}
N. van der Post, E.T. Peters, I.K. Yanson, and J.M. van Ruitenbeek, Phys.
Rev. Lett. {\bf 73}, 2611 (1994).
\bibitem{Urb}
E. Scheer, P. Joyez, M.H. Devoret, D. Esteve, and C. Urbina, Phys. Rev.
Lett. {\bf 78}, 3535 (1997).
\bibitem{Joh}
G. Johansson, E. Bratus', V.S. Shumeiko, and G. Wendin, Physica C {\bf
293}, 77 (1997).
\bibitem{Ye2}
A. Levy Yeyati, J.C. Cuevas, A. Lopez-Davalos, and A. Martin-Rodero, Phys.
Rev. B {\bf 55}, R6317 (1997).
\bibitem{Lan}
R. Landauer, IBM J. Res. Dev. {\bf 1}, 223 (1957). 
\bibitem{But}
M.A. B\"uttiker, Phys. Rev. Lett. {\bf 57}, 1761 (1986). 
\bibitem{Imry}
Y. Imry, in {\em Directions in Condensed Matter Physics} (World Scientific,
Singapore, 1986) p. 102.
\bibitem{WSh2} 
Definition of long SIS junctions was introduced in G. Wendin
and V.S. Shumeiko, Phys. Rev. B {\bf 53}, R6006 (1996).
\bibitem{Dekker}
C. Dekker, Private communication.
\bibitem{Naz}
Yu. V. Nazarov, Physica B {\bf 189}, 57 (1883). 
\bibitem{Delft}
N.C. van der Vaart et al. Phys. Rev. Lett. {\bf 74}, 4702 (1995). 
\bibitem{KBT}
T.M. Klapwijk, G.E. Blonder, and M. Tinkham, Physica B+C {\bf 109-110},
1657 (1982).
\bibitem{Kleiner}
R. Kleiner and P. M\"uller, Phys. Rev. B {\bf 49}, 1327 (1994).
\end{thebibliography}
\end{document}